\documentstyle[12pt]{article}
\input epsf

\begin{document}

\title{Radiative Corrections to Polarized Inelastic Scattering in
Coincidence}

\author{A.V. Afanasev$^{a)}$, I. Akushevich$^{a)}$\footnote{on leave
of absence from the National
Center of Particle and High Energy Physics,
220040 Minsk, Belarus}, G.I. Gakh$^{b)}$, N.P.Merenkov$^{b)}$ }
\date{}
\maketitle
\begin{center}
{\small {\it $^{(a)}$ North Carolina Central University,
Durham, NC 27707, USA \\ and  \\
TJNAF, Newport News, VA 23606, USA\\}}
{\small {\it{$^{(b)}$ NSC "Kharkov Institute of Physics and Technology" \\
}}}
{\small {\it {63108, Akademicheskaya 1, Kharkov, Ukraine}}}
\end{center}
\begin{abstract}
The complete analysis of the model--independent leading radiative
corrections to cross--section and polarization observables in
semi--inclusive deep--inelastic electron-nucleus scattering with
detection of a proton and scattered electron in coincidence has been
performed. The basis of the calculations consists of the Drell--Yan like
representation in electrodynamics for both spin--independent and
spin--dependent parts of the cross--section in terms of the electron
structure functions.  The applications to the
polarization transfer effect from longitudinally polarized electron beam
to detected proton as well as to scattering by the polarized
target are considered.  \end{abstract}
\section{Introduction}

\hspace{0.7cm}

Current experiments at electron accelerators of new generation
reached a new level of precision. Such a precision requires
a new approach to data analysis and inclusion of all possible
systematic uncertainties. One of the important sources of
systematic uncertainties are electromagnetic radiative effects
caused by physics processes in the next orders of perturbation
theory.

The purpose of this paper is developing a unified approach to
computation of radiative effects for inelastic scattering of
polarized electrons in the coincidence setup, namely, when one produced
hadron is detected in coincidence with the scattered electron.
A broad range of measurements falls into the category of coincidence
electron scattering experiments. It includes  deep--inelastic
semi--inclusive leptoproduction of hadrons, $(e,e'h)$, as well as
quasielastic nucleon knock--out processes, $(e,e'N)$. The former
class of experiments gives access to the flavor structure of quark--parton
distributions and fragmentation functions. It is in focus of
experimental programs at CERN, DESY, SLAC and JLab.
Some experiments have already been completed and some are being in
preparation.
%The results on diffractive leptoproduction \cite{HERA}, charge
%asymmetries \cite{SMC}, exclusive $\rho^0$ (or $\phi^0$) muoproduction on
%different nuclei \cite{CERN} and the energy fraction distribution of the
%forward produced hadrons \cite{Fer} in semi--inclusive DIS are published
%during last time.
%An important goal of future $COMPASS$ experiment \cite{COM} is the
%study of semi--inclusive polarized muon scattering off a polarized target.
%$COMPASS$ will be able to study the spin transfer effect in
%semi--inclusive $\Lambda$ production \cite{Lam}.
The
detailed modern review of the activities can be found in \cite{filippone}.
Quasielastic nucleon
knock-out
allows to study single--nucleon properties in nuclear medium
and probe the nuclear wave function \cite{O16,He3}.

The different theoretical aspects of strong interaction in semi--inclusive
DIS were studied in a number of a papers \cite{TA}, \cite{Pol}. The most
direct experimental probe of momentum distribution in nuclei, that is
presently available, is provided by means of the reaction $A(e,e'N)B$ (see
reviews \cite{Rev}). The particular polarization effects in a such type of
reactions on the level of Born approximation with respect to the
electromagnetic interaction have been investigated in Ref.~\cite{SP}.

There are several papers dealing with radiative effects for coincidence
experiments. The lowest order correction was treated in \cite{soroko} using an
approach of covariant cancellation of infrared divergence. Leading log
correction was studied in \cite{Schienbein}
for charm production. At last radiative
correction in quasielastic scattering was recently studied in \cite{templon}.
Different approaches were applied to the calculations and different
approximations were done for that. These calculations adopted some specific
models for structure functions.
Current experimental data do not cover wide enough kinematical ranges, so
extrapolation and interpolation procedures have to be used in calculating
radiative effects. Therefore the model dependence of the results reduces
their generality and as a result, their applicability.
Furthermore higher order effects, which are
important at the current level of experimental accuracies, were not
systematically considered.

The method of the electron structure functions \cite{GL} allows to
treat the observed cross section including both the lowest order
and higher order effects, by the same way. As a result we can
obtain clear and physically transparent formulae for radiative
effects. In this paper we restrict our consideration to leading
accuracy. It allows us to avoid an attraction of any model for the
hadron structure functions and as a result to obtain some general
formulae for quite wide class of the physical processes. In the
case of need the NLO correction to some specific process can be
obtained by standard procedure. Good examples are recent
calculation of LO and NLO correction to polarization observables
in DIS \cite{AAM_DIS} and elastic \cite{AAM_elastic} processes.

In the present paper we consider the model--independent RC to the
cross--section and polarization observables in semi--inclusive
deep--inelastic scattering of the longitudinally polarized
electron off nucleus targets, provided that the target as well as
detected hadron can be polarized. In Sec. 2 we use the electron
structure function approach to calculate RC and derive
the master formulae for the radiatively corrected spin--independent and
spin--dependent parts of the corresponding cross--sections in the
form of the Drell--Yan like representation \cite{DY} in
electrodynamics. The result of this Sec. is suitable for leptonic
variables when the scattered electron is detected too. In Sec. 3
we apply our master formulae to the case when polarization of the
final nucleon is measured.  The account of RC to the
semi--inclusive DIS on the nucleus target with vector polarization
has been performed in Sec. 4. In Sec. 5 we apply our approach to
describe the effects of polarization transfer from the target to
the detected nucleon. These effects includes both double spin
(hadron--hadron) and triple spin (electron--hadron--hadron)
correlations. In Sec. 6 we derive the modification of the master
formulae in the case of hadronic variables (when instead of the
scattered electron the total 4--momentum of the all hadrons is
measured) and consider some applications. Brief discussion of the
expansion of our results for the radiatively corrected polarization
observables beyond the leading-log accuracy is given in
Conclusion.

\section{Master formula}

\hspace{0.7cm} In the recent experiment \cite{EXP} the
polarization transfer to the detected proton in the process with
longitudinally--polarized electron beam $^{^{16}}O(\vec e,e,\vec
p)^{^{15}}N$ has been measured. This reaction is the particular
case of the more general semi--inclusive deep--inelastic polarized
process \begin{equation}\label{process} \vec {e}\ ^-(k_1) + A(p_1)
\longrightarrow e^-(k_2) + \vec {p}\ (p_2) + X.
\end{equation}
In this paper we want to clarify the question how to
calculate the electromagnetic radiative corrections to the cross--section
and polarization observables in the such kind of the process within the
framework of the electron structure function approach.

We will use the following definition of the cross--section of the process
(1) with definite spin orientation of the proton (that is detected in
the final state) in terms of the leptonic and hadronic tensors
\begin{equation}\label{definition of cross--section, 2}
d\sigma =
\frac{\alpha^2}{(2S_A+1)V(2\pi)^3}\frac{L_{\mu\nu}H_{\mu\nu}}{2\hat
q^4}\frac{d^3k_2}{\varepsilon_2}\frac{d^3p_2}{E_2}\ ,
\end{equation}
where $S_A$ is the target spin, $\varepsilon_2 \ (E_2)$ is the energy of
the scattered electron (detected proton) and $\hat q$ is the 4--momentum
of the virtual photon that probes the hadron block.  Hadronic tensor can
be expressed via hadron electromagnetic current $J_{\mu}$ $$H_{\mu\nu} =
\sum_X <p_1|J_{\mu}(\hat q)|p_2,X><X,p_2|J_{\nu}(-\hat
q)|p_1>\delta(P_x^2-M^2_x), \ \ P_x=\hat q +p_1-p_2, $$ where $P_x$ is the
total 4--momentum of the undetected hadron system and $M_x$ is its
invariant mass.

The electron structure function approach yields summation of the
leading--log contributions into the leptonic tensor in all orders of the
perturbation theory. These contributions arise due to radiation of a hard
collinear as well as the soft and virtual photons and electron--positron
pairs by electrons in both, initial and final, states. In the leading
approximation the electron tensor, on the right side of Eq.~(2), can be
written as \cite{M}
\begin{equation}\label{leptonic tensor, 3}
L_{\mu\nu}(k_1,k_2)= \int\int\frac{dx_1dx_2}{x_1x_2^2}
D(x_2,Q^2)[D(x_1,Q^2)\hat Q^{^B}_{\mu\nu}(\hat k_1,\hat
k_2)\ +i\lambda D_{\lambda}(x_1,Q^2)\hat E^{^B}_{\mu\nu}(\hat k_1,\hat
k_2)]\ , \end{equation}
$$Q^2 = -(k_1-k_2)^2\ , \ \hat k_1=x_1k_1, \ \hat
k_2= \frac{k_2}{x_2}\ ,$$ where $D(x,Q^2)$ is the structure
function that describes radiation of an unpolarized electron, and
$D_{\lambda} (x,Q^2)$ -- of longitudinally--polarized one. On the
level of the next--to--leading accuracy these functions differ
already in the first order of the perturbation theory, but in the
framework of the used here leading one, in the second order only.
The corresponding difference is caused by leading contribution into
$D$--function due to $e^+e^-$--pair production in the singlet channel
(effect of the final--electron identity), which is different for
unpolarized and longitudinally polarized electron and read
\cite{M} (KMS), \cite{KB} $$D^{^S} = \Bigl(\frac{\alpha
L}{2\pi}\Bigr)^2\bigl[\frac{2(1-x^3)}{3x}
+\frac{1-x}{2}+(1+x)\ln{x}\bigr]\ , \ \ L=\frac{Q^2}{m_e^2}\ ,$$
$$D^{^S}_{\lambda} = \Bigl(\frac{\alpha
L}{2\pi}\Bigr)^2\bigl[\frac{5(1-x)}{2}+(1+x)\ln{x}\bigr]\ , $$
where $m_e$ is the electron mass.

The accounting of the singlet channel contribution leads usually to very
small effects (of the order $10^{-4}$) because, as one can see, terms into
brackets trend to compensate each other (see, for example, \cite{JK}).
Below we will not distinguish between $D$ and $D_{\lambda}$, which
corresponds to the accounting of the nonsinglet channel contribution only
(for the corresponding $D$--functions see \cite{KB,JK}).
Such approximation allows to write compact formulae for the radiatively
corrected cross--sections. We will also omit quantity $Q^2$ from arguments
of the $D$--functions.

The quantity $\lambda,$ on the right side of Eq.~(3), is the degree of
longitudinal polarization of the electron beam. The limits of the
integration will be defined below.  The representation (3) follows from
the quasi--real electron approximation \cite{B}. The physical sense of
variables $x_1$ and $x_2$ is as follows:  $1-x_1$ is the energy fraction
of all collinear photons and $e^+e^-$--pairs, radiated by the initial
electron, respect to its energy, $ 1-x_1=\omega/\varepsilon_1,$ and
quantity $(1-x_2)/x_2$ is the same for the scattered electron.

In the Born approximation
\begin{equation}\label{Born L, 4}
Q^{^B}_{\mu\nu}(k_1,k_2)=q^2 g_{\mu\nu} +
2(k_1k_2)_{\mu\nu}\ , \ E^{^B}_{\mu\nu}(k_1,k_2)=2(\mu\nu k_1 k_2)\ , \ \
(\mu\nu k_1 k_2)= \epsilon_{\mu\nu\rho\sigma}k_{1\rho}k_{2\sigma} \ ,
\end{equation}
$$ (k_1k_2)_{\mu\nu}=k_{1\mu}k_{2\nu}+k_{1\nu}k_{2\mu}, \ \ q = k_1-k_2\
.$$

The hadronic tensor, on the right side of Eq.~(2), in general case depends
on 4--momenta $p_1,\ p_2$, 4--momentum of the virtual photon $\hat q = \hat
k_1-\hat k_2$, and 4--vector of the hadron spin
$S$ that satisfies conditions: $S^2 =-1, \ (Sp_2)=0.$ For example, in the
case under consideration $$H_{\mu\nu} =
H_{\mu\nu}^{(u)}+H_{\mu\nu}^{(p)}\ ,$$ \begin{equation}\label{H,5}
H_{\mu\nu}^{(u)}= h_1\tilde g_{\mu\nu}+h_2\tilde p_{1\mu}\tilde p_{1\nu}
+h_3\tilde p_{2\mu}\tilde p_{2\nu} + h_4(\tilde p_1\tilde p_2)_{\mu\nu}
+ih_5[\tilde p_1\tilde p_2]_{\mu\nu}\ , \end{equation}
$$H_{\mu\nu}^{(p)}=(Sp_1)\big[h_6(\tilde
p_1N)_{\mu\nu}+ih_7[\tilde p_1N]_ {\mu\nu}+h_8(\tilde
p_2N)_{\mu\nu}+ih_9[\tilde p_2N]_{\mu\nu}\big]+(S\hat
q)\bigl[h_{10}(\tilde p_1N)_{\mu\nu}+$$
$$ih_{11}[\tilde p_1N]_{\mu\nu}+
h_{12}(\tilde p_2N)_{\mu\nu}+ih_{13}[\tilde p_2N]_{\mu\nu}\big] +
(SN)\big[h_{14}\tilde g_{\mu\nu}+h_{15}\tilde p_{1\mu}\tilde
p_{1\nu}+h_{16} \tilde p_{2\mu}\tilde p_{2\nu} +$$
\begin{equation}\label{H(p),6}
h_{17}(\tilde p_1\tilde p_2)_{\mu\nu}+ih_{18}[\tilde p_1\tilde p_2]_
{\mu\nu}]\ , \ \ N_{\mu} = \epsilon_{\mu\nu\rho\sigma}p_{1\nu}p_{2\rho}
\hat q_{\sigma} = (\mu
p_1p_2\hat q)\ , \ \ [ab]_{\mu\nu}=a_{\mu}b_{\nu}-a_{\nu}b_{\mu}\ ,
\end{equation}
$$\tilde g_{\mu\nu} = g_{\mu\nu}-\frac{\hat q_{\mu}\hat q_{\nu}}{\hat
q^2}\ , \ \tilde p_{i\mu}=p_{i\mu}-\frac{(\hat qp_i)\hat q_{\mu}}{\hat
q^2}\ , \ \ i=1,2 \ ,$$
where $h_i \ (i=1-18)$ are the hadron
semi--inclusive structure functions which depend in general on four
invariants. These invariants can be taken as $\hat q^2,\ (\hat q p_1),\
(\hat q p_2)$ and $(p_1p_2).$

The j--component of the proton polarization $P^{^j}$, that could be
measured in experiment, is defined as the ratio of the spin--dependent
part of the cross--section (2) (which is caused by contraction of the
leptonic tensor with the spin--dependent part of the
hadronic one $H^{(p)}_{\mu\nu},$ with the given j--component of the proton
spin) to the spin--independent one (which is caused by contraction of
$L_{\mu\nu}$ with $H^{(u)}_{\mu\nu}$)
\begin{equation}\label{Polarization,7}
P^{^j}=\frac{d\sigma^{(p)}(\lambda,S^{^j},k_1,k_2,p_1,p_2)}
{d\sigma^{(u)}(\lambda,k_1,k_2,p_1,p_2)}\ .  \end{equation}
Note that $P^{^j}$ is non--zero even if $\lambda =0$ (the case of
unpolarized electron beam) due to non--zero single--spin correlations in
semi--inclusive processes.

In the process (1) one can
measure, in principle, three independent components:  $P^{^l}$
(longitudinal), $P^{^t}$ (transverse) and $P^{^n}$ (normal), which
could be taken respect to definite physical directions and planes created
by 3--momenta of the particles participating in the process. If any
additional particle (photons and $e^+e^-$-pairs), radiated by electrons
with 4--momenta $k_1$ and $k_2,$ is not detected, there are three
independent directions: along $\vec p_2,\ \vec k_1$ and $\vec k_2.$ In
this case any components of the proton polarization as well as the
corresponding proton spin components $S^{^j}$ will be defined for the Born
kinematics and their directions are not affected by radiation.

Combining formulae for the cross--section (2), the definitions of the
lepton (3,4) and hadron (5,6) tensors and taking into account the
last discussions, we can write the following representation for the
cross--section of the process (1)
\begin{equation}\label{representation,8}
\varepsilon_2E_2\frac{d\sigma(\lambda,S^{^j},k_1,k_2,p_1,p_2)}{d^3k_2d^3p_2}
= \int\int\frac{dx_1dx_2}{x_2^2}
D(x_1)D(x_2)
\hat\varepsilon_2E_2\frac{d\sigma^{^B}(\lambda,S^{^j},\hat k_1,\hat
k_2,p_1,p_2)} {d^3\hat k_2d^3p_2}\ , \end{equation} where $j=l,\
t,\ n.$ The factor $1/x_1$ that enters into definition of $L_{\mu\nu}$
is absorbed into flow in the reduced Born cross--section that equals by
definition (see Eq.~(2))
$$\hat\varepsilon_2E_2
\frac{d\sigma^B(\lambda,S^{^j},\hat k_1,\hat k_2,p_1,p_2)}{d^3\hat
k_2d^3p_2}=
\frac{\alpha^2}{(2S_A+1)\hat V(2\pi)^3}\frac{L_{\mu\nu}^B(\hat k_1,\hat
k_2,\lambda) H_{\mu\nu}(S^j,\hat q,p_1,p_2)}{2\hat q^4}
 \ , $$ where $\hat V = x_1V.$ With the chosen accuracy the representation
 (8) is valid for both spin--dependent $(d\sigma^{(p)})$ and
 spin--independent $(d\sigma^{(u)})$ parts of the cross--section.

In theoretical calculations it is useful often to parameterize the proton
spin 4--vector, which enters in definition of the hadron tensor, in terms
of the particle 4--momenta \cite{P}. In considered case we have four
4--momenta to express any component of the proton spin $S^{^j}$ in a such
way that
\begin{equation}\label{9}
 S^{^j} = S^{^j}(k_1,k_2,p_1,p_2)\ .
\end{equation}
Let us imagine for a moment that chosen parameterization on the right
side of Eq.~(9) is stabilized relative substitution
$$k_1 \rightarrow \hat k_1\ , \ \ k_2 \rightarrow \hat k_2 \ , \ \
S^{j_s}(k_1,k_2,p_1,p_2) = S^{j_s}(\hat k_1,\hat k_2,p_1,p_2)\ .$$
(Further we will label such stabilized parameterizations by the index with
small letter). In this case we can write the
Born cross--section under integral sign on the right side of Eq.~(8) in
the form \begin{equation}\label{10}
\hat\varepsilon_2E_2\frac{d\sigma^{^B}(\lambda,S^{j},\hat k_1,\hat
k_2,p_1,p_2)} {d^3\hat k_2d^3p_2} =
\hat\varepsilon_2E_2\frac{d\sigma^{^B}_{j}(\lambda, \hat k_1,\hat
k_2,p_1,p_2)}{d^3\hat k_2d^3p_2}\ .  \end{equation}

If the proton spin $S^{^J}$ is unstable under above substitution
(in this case we will use the index with capital letter) it can be
expressed always in terms of stabilized one by means of orthogonal
matrix \begin{equation}\label{11} S^{^J}(k_1,k_2,p_1,p_2) =
A_{Jj}(k_1,k_2,p_1,p_2)S^{^j}(\hat k_1,\hat k_2,p_1,p_2)\ , \ \
A_{Jj} = - S^{^J}S^{^j}\ .
\end{equation}

Using the last formula and taking into account that in the
considered class of the processes the hadron tensor depends
linearly on the proton spin, we can write the master
representation for the spin--dependent part $(d\sigma^{(p)})$ of
the cross--section of the process (1) for arbitrary orientation of
the proton spin in the following form
\begin{equation}\label{master formula 12}
\varepsilon_2E_2\frac{d\sigma(\lambda,S^{^J},k_1,k_2,p_1,p_2)}{d^3k_2d^3p_2}
= A_{Jj}\int\int\frac{dx_1dx_2}{x_2^2} D(x_1)D(x_2)
\hat\varepsilon_2E_2\frac{d\sigma^{^B}_j(\lambda,\hat k_1,\hat
k_2,p_1,p_2)} {d^3\hat k_2d^3p_2}\ , \end{equation} where we bear
in mind the summation over index $j = l,t,n.$

This representation is the electrodynamical analogue of the well
known in QCD Drell--Yan formula \cite{DY}, that was applied
earlier to calculate the electromagnetic radiative corrections to
the total cross--section of the electron--positron annihilation
into hadrons \cite{KB}, to small-angle Bhabha scattering
cross--section at LEP1 \cite{JK}, to unpolarized \cite{KMF} and
polarized deep--inelastic cross--sections \cite{AAM_DIS}, and to
polarized elastic electron--proton scattering
\cite{AAM_elastic}. In the next Section we will show how this
representation can be used to describe the leading radiative
corrections in polarized semi--inclusive deep--inelastic events. It
is obvious that in the framework of the leading accuracy one needs
to find the adequate parameterizations of the proton spin
4--vector, to calculate the elements of the orthogonal matrix
$A_{Jj}$, derive the spin--independent and spin--dependent parts
of the Born cross--section for given parameterization $S^{^j},$ and
determine the limits of the integration over $x_1$ and $x_2$ in the
master formula (12).

\section{Analysis of semi--inclusive deep--inelastic events with
polarization transfer}

\hspace{0.7cm}

Let us begin with the parameterizations of the proton spin 4--vector in
process (1). To describe this process we will use the following set of
invariant variables
\begin{equation}\label{13, variables}
z=\frac{2p_1p_2}{V}\ , \ z_{1,2} = \frac{2k_{1,2}p_2}{V}\ , \
y=\frac{2p_1(k_1-k_2)}{V}\ , x=\frac{-q^2}{2p_1q}\ , V=2p_1k_1, \
q=k_1-k_2\ .
\end{equation}

It is physically justified to determine the longitudinal component
of the proton spin along direction of $-\vec p_1$ as seen from the
rest frame of the detected proton.  This direction does not
affected by the lepton collinear radiation and the corresponding
parameterization has a form
\begin{equation}\label{14,|| spin}
S_{\mu}^{^l}=\frac{zp_{2\mu}-2\tau_2p_{1\mu}}{m\sqrt{z^2-4\tau_1
\tau_2}}\ , \ \ \tau_1=\frac{M^2}{V}\ , \ \ \tau_2=\frac{m^2}{V}\ ,
\end{equation}
where $M(m)$ is the mass of the target nucleus (detected proton). It is
easy to verify that in the rest frame of proton $(p_2=(m,0))$ this
longitudinal component equals to $(0,-\vec n_1),$ where $\vec n_1 = \vec
p_1/|\vec p_1|,$ and in the lab. system $(p_1=(M,\ 0))$ it equals to
$(|\vec p_2|, \ E_2\vec n_2)/m,$ where $\vec n_2$ is the unit vector in
direction of the detected proton 3--momentum.

For the fixed longitudinal component we have a few possibilities
to determine the transverse and normal ones. First, take the
transverse component in the plane $(\vec k_1,\ \vec p_2)$ and the
normal component in the plane that is perpendicular to it.
Orientations of these planes do not change during substitution
$\vec k_1\rightarrow \hat{\vec k_1},$ therefore in this case we
have \begin{equation}\label{15} S^{^t}_{\mu} =
\frac{(z^2-4\tau_1\tau_2)k_{1\mu}+(2z_1\tau_1-z)p_{2\mu}+
(2\tau_2-zz_1)p_{1\mu}}{\sqrt{V(z^2-4\tau_1\tau_2)[1]}}\ , \ \
S^{^n}_{\mu} = \frac{2(\mu k_1p_1p_2)}{\sqrt{V^3[1]}}\ ,
\end{equation}
$$ [1] = zz_1-\tau_2-z_1^2\tau_1 \ , \ \ (S^{^j}S^{^i})=
-\delta_{ji}\ . $$
By full analogy with above procedure we can determine other stabilized set
of transverse and normal components relative to the plane $(\vec k_2,\
\vec p_2)$ \begin{equation}\label{16,2} \tilde S^{^t}_{\mu} =
\frac{(z^2-4\tau_1\tau_2)k_{2\mu}+(2z_2\tau_1-z(1-y))p_{2\mu}+
(2\tau_2(1-y)-zz_2)p_{1\mu}}{\sqrt{V(z^2-4\tau_1\tau_2)[2]}}\ ,
\end{equation}
$$\tilde S^{^n}_{\mu} = \frac{2(\mu k_2p_1p_2)}{\sqrt{V^3[2]}}\ ,\ \
[2] = zz_2(1-y)-\tau_2(1-y)^2-z_2^2\tau_1 \ .$$

The sets (15) and (16) represent the complete list of the
stabilized parameterizations of the proton spin components on the
condition that the longitudinal component is chosen according to
Eq.~(14). There are a lot of unstable parameterizations because we
can take them relative to arbitrary plane $(a\vec k_1 + b\vec
k_2,\ \vec p_2)$ with arbitrary numbers $a$ and $b.$ In further we
will consider the physically favorable set with $a=-b=1$ only.
The corresponding transverse and normal components read
\begin{equation}\label{17, q-set} S^{^T}_{\mu} =
\frac{(z^2-4\tau_1\tau_2)q_{\mu}+(2(z_1-z_2)\tau_1-zy)p_{2\mu}+
(2y\tau_2-z(z_1-z_2))p_{1\mu}}{\sqrt{V(z^2-4\tau_1\tau_2)[q]}}\ ,
\end{equation}
$$S^{^N}_{\mu} = \frac{2(\mu qp_1p_2)}{\sqrt{V^3[q]}}\ , \ \
[q] = zy(z_1-z_2) +xy(z^2-4\tau_1\tau_2) -(z_1-z_2)^2\tau_1
-y^2\tau_2\ . $$

Let us consider now the relation between stabilized (for
definiteness we will work with the set (15)) set and unstable one.
It is obvious that this relation can be written as follows
\begin{equation}\label{18,connection spin}
S^{^N} = \cos{\theta}S^{^n} + \sin{\theta}S^{^t} \ , \
S^{^T} = - \sin{\theta}S^{^n} + \cos{\theta}S^{^t} \ ,
\end{equation}
where
$$\cos{\theta} = -(S^{^N}S^{^n}) = -(S^{^T}S^{^t})
=\frac{z(z_1(1+y)-z_2)+xy(z^2-4\tau_1\tau_2)-2z_1(z_1-z_2)
\tau_1-2y\tau_2}{2\sqrt{[1][q]}}\ , $$
$$\sin{\theta} = -(S^{^N}S^{^t}) = (S^{^T}S^{^n}) =
\frac{\eta}{2}\sqrt{\frac{(z^2-4\tau_1\tau_2)}{[1][q]}}\ , $$
$$\eta=sign[(p_1p_2k_1k_2)]\sqrt{\frac{16}{V^4}(p_1p_2k_1k_2)^2}, \
(p_1p_2k_1k_2)=\epsilon_{\mu\nu\rho\sigma}p_{1\mu}p_{2\nu}k_{1\rho}
k_{2\sigma}, $$
$$ \frac{16(p_1p_2k_1k_2)^2}{V^4}=x^2y^2(4\tau_1\tau_2-z^2)+
2xy[z(z_2+z_1(1-y))-2z_1z_2\tau_1-2(1-y)\tau_2]-(z_2-z_1(1-y))^2\ .$$
One can verify that the necessary condition $\cos^2\theta+\sin^2\theta=1$
is satisfied.

Now we can write down the spin--independent (we bear in mind
that it means independent on the proton spin only) and spin--dependent
parts of the cross--section of the process (1) as
\begin{equation}\label{19, unpolarized}
\varepsilon_2E_2\frac{d\sigma_{(u),L}}{d^3k_2d^3p_2}=
\int\int\frac{dx_1dx_2}{x_2^2}D(x_1)
D(x_2)\hat\varepsilon_2E_2\frac{d\hat\sigma_{(u),l}^B}
{d^3\hat k_2d^3p_2} \ ,
\end{equation}
\begin{equation}\label{20, polarized,N}
\varepsilon_2E_2\frac{d\sigma_{N}}{d^3k_2d^3p_2}=
\int\int\frac{dx_1dx_2}{x_2^2}D(x_1)
D(x_2)\hat\varepsilon_2E_2\Bigl[\cos\theta\frac{d\hat\sigma_{n}^B}
{d^3\hat k_2d^3p_2} + \sin\theta\frac{d\hat\sigma_{t}^B}{d^3\hat
k_2d^3p_2}\Bigl] , \end{equation}
\begin{equation}\label{21, polarized,bot}
\varepsilon_2E_2\frac{d\sigma_{T}}{d^3k_2d^3p_2}=
\int\int\frac{dx_1dx_2}{x_2^2}D(x_1)
D(x_2)\hat\varepsilon_2E_2\Bigl[-\sin\theta\frac{d\hat\sigma_{n}^B}
{d^3\hat k_2d^3p_2} + \cos\theta\frac{d\hat\sigma_{t}^B}{d^3\hat
k_2d^3p_2}\Bigl] , \end{equation}
where $d\hat\sigma^B,$ with any low index, denotes the corresponding Born
cross--section given at $shifted$ values of $k_{1,2}\rightarrow\hat
k_{1,2}.$ The corresponding $shifted$ dimensionless variables, introduced
by relation (13), read \begin{equation}\label{22,shifted} \hat x
=\frac{x_1xy}{x_1x_2+y-1}\ , \ \hat y =\frac{x_1x_2+y-1}{x_1x_2} \ , \
\hat V=x_1V, \ \hat z = \frac{z}{x_1}, \ \hat z_1 = z_1, \ \hat
z_2=\frac{z_2}{x_1x_2}\ .  \end{equation} Eqs.(19)--(21) are the
straightforward consequences of the master representation (12). In order
to obtain $d\sigma_n$ and $d\sigma_t$ on the left side of Eqs.~(20) and
(21) we have to take obviously $\cos\theta=1, \ \sin\theta=0.$

Now we must derive the Born cross--sections which enter on the right sides
of Eqs.~(19)--(21). The spin--independent part of the cross--section for
longitudinally- polarized electron beam (with degree $\lambda$) is
expressed in terms of the hadron structure functions $h_1-h_5$ as
\begin{equation}\label{23, unpolarized Born}
\varepsilon_2E_2\frac{d\sigma^B_{(u)}}{d^3k_2d^3p_2}=\frac{\alpha^2V}
{2(2S_A+1)(2\pi)^3q^4}H_1\ ,
\end{equation}
$$H_1=-\frac{2xy}{V}h_1+(1-y-xy\tau_1)h_2+(z_1z_2-xy\tau_2)h_3+
(z_2+z_1(1-y)-xyz)h_4-\lambda\eta h_5\ .$$
Note that the phase space of the detected proton also can be expressed in
terms of invariant variables (13)
\begin{equation}\label{24}
\frac{d^3p_2}{E_2}=\frac{V}{2|\eta|}dz_1dz_2dz \ .
\end{equation}

If the proton spin is directed along $S^{^l}$ then the
spin--dependent part of the Born cross-section reads
\begin{equation}\label{25}
\varepsilon_2E_2\frac{d\sigma^B_{l}}{d^3k_2d^3p_2}=-
\frac{\alpha^2V^3\eta\sqrt{z^2-4\tau_1\tau_2}}{8(2S_A+1)m(2\pi)^3q^4}
\Bigl[H_2 +
\frac{[z(z_1-z_2)-2y\tau_2]}{z^2-4\tau_1\tau_2}H_3\Bigr]\ ,
\end{equation}
$$H_2=(2-y)h_6+(z_1+z_2)h_8+ \frac{\lambda}{\eta}
(\eta_1h_7+\eta_2h_9)\ , $$
$$H_3=(2-y)h_{10}+(z_1+z_2)h_{12}+\frac{\lambda}{\eta}
(\eta_1h_{11}+\eta_2h_{13}) \ ,$$
$$\eta_1=y[z_2-z_1(1-y)-xz(2-y)+2x(z_1+z_2)\tau_1]\ , $$
$$\eta_2=(z_1-z_2)(z_2-z_1(1-y))+xyz(z_1+z_2)-2xy(2-y)\tau_2\ . $$

In the case of transverse orientation of the proton spin (along
$S^{^t}$) we have
\begin{equation}\label{26, bots}
\varepsilon_2E_2\frac{d\sigma^B_{t}}{d^3k_2d^3p_2}=
\frac{\alpha^2V^2\eta}{8(2S_A+1)(2\pi)^3q^4}\sqrt{\frac{V}{z^2-4\tau_1\tau_2}}
\Bigl[\psi H_3-\frac{z^2-4\tau_1\tau_2}{\sqrt{[1]}}H_4\Bigr]\ ,
\end{equation}
$$\psi =
\frac{xy(z^2-4\tau_1\tau_2)+(z-2z_1\tau_1)(z_1-z_2)+(zz_1-2\tau_2)y}
{\sqrt{[1]}} = 2\sqrt{[q]}\cos\theta\ ,  $$
where $H_4$ can be obtained from $H_1$ by means of simple replacement
$h_i\rightarrow h_{i+13}.$

At last, for the normal orientation of the proton spin (along $S^{^n}$) the
spin--dependent part of the cross--section of the process (1) reads
\begin{equation}\label{27, Ns}
\varepsilon_2E_2\frac{d\sigma^B_{n}}{d^3k_2d^3p_2}=
\frac{\alpha^2V^2\sqrt{V}}{8(2S_A+1)(2\pi)^3q^4}
\Bigl[-\frac{\eta^2}{\sqrt{[1]}}H_3-
\psi H_4\Bigr]\ .
\end{equation}

We have to determine also the limits of integration over variables
$x_1$ and $x_2$ in the master representation (12). They can be obtained
from the condition that the semi--inclusive deep--inelastic
process takes place. For an electron--proton scattering it is possible
on the condition that the hadron state consists,
at least, of a proton and a pion. This leads to inequality
\begin{equation}\label{28,limits}
x_1x_2+y-1-x_1xy\geq x_2\delta\ ,\ \ \delta = \frac{(m+m_{\pi})^2
-m^2}{V}\ ,
\end{equation}
where $m_{\pi}$ is the pion mass. This inequality yields for the limits
\begin{equation}\label{29}
1>x_2>\frac{1-y+xyx_1}{x_1-\delta}\ , \ \ 1>x_1>\frac{1+\delta-y}{1-xy}\ .
\end{equation}
For the electron--nucleus scattering process (1) that is considered here
we must change the pion mass, in definition of $\delta,$ by the bound
energy of the ejected proton in a given nucleus.

It is interesting to note that in the case, when the polarizations
of the final proton are measured relative to stabilized
orientations, the corresponding Born values and the leading
radiative corrections to them are expressed in terms of the same
hadron structure functions. The situation changes radically if one
measures polarizations relative to the unstable orientations. In
this case the contributions to the polarizations, caused by the
radiative corrections due to hard collinear radiation, are
expressed in terms of another sets of hadron structure functions
as compared with the Born polarizations. To give this fact more
transparent, we write down the spin--dependent part of the Born
cross--section for the orientations of the proton spin along
$S^{^N}$ and $S^{^T}$ \begin{equation}\label{30}
\varepsilon_2E_2\frac{d\sigma^B_{T}}{d^3k_2d^3p_2}=
\frac{\alpha^2V^2\eta}{4(2S_A+1)(2\pi)^3q^4}\sqrt{\frac{V[q]}
{z^2-4\tau_1\tau_2}}H_3\ ,
\end{equation}
\begin{equation}\label{31}
\varepsilon_2E_2\frac{d\sigma^B_{N}}{d^3k_2d^3p_2}=
-\frac{\alpha^2V^2\sqrt{V[q]}}{4(2S_A+1)(2\pi)^3q^4}H_4\ .
\end{equation}
These formulae can be derived from Eqs.~(20) and (21) if to take
$D(x_{i})$--functions in form of $\delta$--function, which corresponds to
the radiationless process (or to the Born approximation).

\section{Semi--inclusive deep--inelastic scattering on polarized target}

\hspace{0.7cm}

In this section we will apply the master representation to the analysis of
polarized phenomena in semi--inclusive deep--inelastic scattering of
polarized nucleus
\begin{equation}\label{32}
\vec {e}^-(k_1) + \vec A(p_1)\rightarrow e^-(k_2) + H(p_2) +X \ ,
\end{equation}
where $H$ is arbitrary hadron and nucleus $A$ has definite vector
polarization $P.$ In this case the leptonic tensor is as before
(see Eqs.~(3) and (4)), and the hadronic tensor has the same
structure as defined by Eqs.~(5) and (6), where one needs to use
polarization of the nucleus $P$ instead of the proton spin $S$ and
write $(Pp_2)$ instead of $(Sp_1).$ Besides, we will use the
notation $g_1-g_{18}$ for the corresponding hadron structure
functions.

Usually when studying the polarization phenomena the various
asymmetries are measured and to find them it is necessary to know the
polarization--independent and polarization--dependent parts of the
cross--section at different orientations of the target polarization.
Therefore, the corresponding analysis can be performed in the same manner
as it was done in Section 2.

Let us, at first, define the parameterizations of the nucleus polarization
4--vector in terms of 4--momenta. As a stabilized set we can choose
longitudinal and transverse components as given in Ref.~\cite{AAM_DIS}
\begin{equation}\label{33}
P^{^l}_{\mu} = \frac{2\tau_1k_{1\mu}-p_{1\mu}}{M}\ , \
P^{^t}_{\mu}=\frac{k_{2\mu}-(1-y-2xy\tau_1)k_{1\mu}
-xyp_{1\mu}}{\sqrt{Vxy(1-y-xy\tau_1)}}\ ,
\end{equation}
and for normal component we use
\begin{equation}\label{34}
P^{^n}_{\mu}=\frac{2(\mu k_1k_2p_1)}{\sqrt{V^3xy(1-y-xy\tau_1)}}\ .
\end{equation}
It is easy to verify that parameterizations (33), (34) are not changed
at the substitution $k_{1,2}\rightarrow\hat k_{1,2}.$ In lab. system this
set corresponds to direction of the longitudinal polarization along
$\vec k_1,$ the transverse polarization is in the plane $(\vec k_1,\ \vec
k_2)$ and the normal one is in the plane, that is perpendicular to $(\vec
k_1,\ \vec k_2)$ plane.

Another set of the polarizations can be chosen in a such way that
longitudinal component will be along $\vec q$--direction in lab. system
and the transverse one is in the plane $(\vec q,\ \vec k_1).$ In this case
the normal component coincides with (34) and \begin{equation}\label{35}
P^{^L}_{\mu}= \frac{2\tau_1(k_{1\mu}-k_{2\mu})-yp_{1\mu}}{M
\sqrt{y^2+4xy\tau_1}}\ , \
P^{^T}_{\mu}=\frac{(1+2x\tau_1)k_{2\mu}-(1-y-2x\tau_1)k_{1\mu}
-x(2-y)p_{1\mu}}{\sqrt{Vx(1-y-xy\tau_1)(y+4x\tau_1)}}\ .
\end{equation}

The sets (35) and (33) are transformed one to other by orthogonal matrix
$$P^{^L}=\cos\theta_1P^{^l} +\sin\theta_1P^{^t}\
, \ P^{^T}=-\sin\theta_1P^{^l} +\cos\theta_1P^{^t}\ ,
$$ \begin{equation}\label{36}
\cos\theta_1=\frac{y(1+2x\tau_1)}{\sqrt{y(y+4x\tau_1)}}\ , \ \sin\theta_1
= -2\sqrt{\frac{x\tau_1(1-y-xy\tau_1)}{y+4x\tau_1}}\ .
\end{equation}

The master equation (12) can be applied to the
polarization--independent part of the cross--section (32) as well
as to the polarization--dependent one.  Therefore, we have to
derive the Born cross--section for the stabilized set.  The simple
calculation gives
\begin{equation}\label{37, unpolarized Born}
\varepsilon_2E_2\frac{d\sigma^B_{(u)}}{d^3k_2d^3p_2}=\frac{\alpha^2V}
{(2S_A+1)(2\pi)^3q^4}G_1\ .
\end{equation}
%$$G_1=\bigl[-\frac{2xy}{V}g_1+(1-y-xy\tau_1)g_2+(z_1z_2-xy\tau_2)g_3+
%(z_2+z_1(1-y)-xyz)g_4-\lambda\eta g_5\bigr]\ .$$
Note that numerical coefficient in front of $G_1$ is twice as much as
compared with that on the right side of Eq.~(23) in front of $H_1.$ The
reason is that in this case we do not fix the spin state of the final
hadron $H.$

The polarization--dependent part of the cross--section for the longitudinal
stabilized polarization reads \begin{equation}\label{38}
\varepsilon_2E_2\frac{d\sigma^B_{l}}{d^3k_2d^3p_2}=
-\frac{\alpha^2V^3\eta}{4(2S_A+1)M(2\pi)^3q^4}\bigl[(2\tau_1z_1-z)G_2
-y(1+2x\tau_1)G_3 +2\tau_1G_4\bigr]\ ,
\end{equation}
%$$G_2=(2-y)g_6+(z_1+z_2)g_8+\frac{\lambda}{\eta}(\eta_1g_7
%+\eta_2g_9)\ , $$
%$$G_3=(2-y)g_{10}+(z_1+z_2)g_{12}+\frac{\lambda}{\eta}(\eta_1g_{11}
%+\eta_2g_{13})\ , $$
where the functions $G_i, \ i=1-4,$ can be derived from $H_i$ by
replacement the hadron structure functions $g_j$ instead of $h_j.$

The corresponding part of the cross--section in the case of the
transverse polarization can be written as follows
\begin{equation}\label{39}
\varepsilon_2E_2\frac{d\sigma^B_{t}}{d^3k_2d^3p_2}=
-\frac{\alpha^2V^2\eta\sqrt{Vxy(1-y-xy\tau_1)}}{4(2S_A+1)(2\pi)^3q^4}
\bigl[\frac{z_2-xyz-z_1(1-y-2xy\tau_1)}{xy(1-y-xy\tau_1)}G_2+
\end{equation}
$$2G_3 +\frac{1+2x\tau_1}{x(1-y-xy\tau_1)}G_4\bigr]\ .$$

For the normal polarization the spin--dependent part of the cross--section
is
\begin{equation}\label{40}
\varepsilon_2E_2\frac{d\sigma^B_{n}}{d^3k_2d^3p_2}=
\frac{\alpha^2V^2}{4(2S_A+1)(2\pi)^3q^4}\sqrt{\frac{V}{xy(1-y-xy\tau_1)}}
\bigl[\eta^2 G_2-y\bigl(z_2(1+2x\tau_1)-
\end{equation}
$$z_1(1-y-2x\tau_1)-xz(2-y)\bigr)G_4\bigr]\ .$$

The application of the master representation (12) leads to
following expressions for radiatively corrected (with the leading
accuracy) contributions to the cross--section of the process (32)
\begin{equation}\label{41, unpolarized,N}
\varepsilon_2E_2\frac{d\sigma_{(u),N}}{d^3k_2d^3p_2}=
\int\int\frac{dx_1dx_2}{x_2^2}D(x_1)
D(x_2)\hat\varepsilon_2E_2\frac{d\hat\sigma_{(u),n}^B}
{d^3\hat k_2d^3p_2} \ ,
\end{equation}
\begin{equation}\label{42, polarized,||}
\varepsilon_2E_2\frac{d\sigma_{L}}{d^3k_2d^3p_2}=
\int\int\frac{dx_1dx_2}{x_2^2}D(x_1)
D(x_2)\hat\varepsilon_2E_2\Bigl[\cos\theta_1\frac{d\hat\sigma_{l}
^B}{d^3\hat k_2d^3p_2} + \sin\theta_1\frac{d\hat\sigma_{t}^B}{d^3\hat
k_2d^3p_2}\Bigl] , \end{equation}
\begin{equation}\label{43, polarized,bot}
\varepsilon_2E_2\frac{d\sigma_{T}}{d^3k_2d^3p_2}=
\int\int\frac{dx_1dx_2}{x_2^2}D(x_1)
D(x_2)\hat\varepsilon_2E_2\Bigl[-\sin\theta_1\frac{d\hat\sigma_{l}
^B}{d^3\hat k_2d^3p_2} + \cos\theta_1\frac{d\hat\sigma_{t}^B}{d^3\hat
k_2d^3p_2}\Bigl] . \end{equation}

Let us write also the cross--sections, on the left sides of Eqs.(42) and
(43), in the Born approximation
\begin{equation}\label{44}
\varepsilon_2E_2\frac{d\sigma_{L}^B}{d^3k_2d^3p_2}=
\frac{\alpha^2V^3\eta}{4(2S_A+1)(2\pi)^3Mq^4}\Bigl[\frac{yz-2(z_1-z_2)\tau_1}
{\sqrt{y(y+4x\tau_1)}}G_2
+\sqrt{y(y+4x\tau_1)}G_3\bigr]\ ,
\end{equation}
\begin{equation}\label{45}
\varepsilon_2E_2\frac{d\sigma_{T}^B}{d^3k_2d^3p_2}=
\frac{\alpha^2V^2\eta\sqrt{V}}{4(2S_A+1)(2\pi)^3q^4}\Bigl[-\sqrt{\frac{y+4x\tau_1}
{x(1-y-xy\tau_1)}}G_4+
\end{equation}
$$\frac{xz(2-y)-z_2+z_1(1-y)-2x\tau_1(z_1+z_2)}
{\sqrt{x(y+4x\tau_1)(1-y-xy\tau_1)}}G_2\Bigr]\ .$$
As one can see, the polarization--dependent parts of the Born
cross--section consist of less number of the hadron structure functions
as compared with radiatively corrected ones.

We can also use the 4--vector $p_2$ to parameterize the nucleus polarization
4--vector. If to choose the longitudinal polarization along $\vec p_2$ in
the lab. system, then the stabilized set may be defined with respect to
the plane $(\vec k_1,\ \vec p_2)$ and unstable one with respect to
the plane $(\vec q, \ \vec p_2) $ as in Section 2, and the corresponding
calculations are very close to given there. But parameterizations, used in
this Section, look more physically and they can be used also to describe
the polarization phenomena in inclusive deep--inelastic events.

\section{Polarization transfer from target to detected proton}

Let us consider effects of the polarization transfer from the vector
polarized target to detected proton in the process
\begin{equation}\label{46} \vec e^{\ -}(k_1) + \vec A(p_1)\rightarrow
e^-(k_2) + \vec p(p_2) +X\ \end{equation} for the case of longitudinally
polarized electron beam and vector polarization of the target.

The general form of the hadronic tensor in this case reads
\begin{equation}\label{47}
H_{\mu\nu}=H^{^{(u)}}_{\mu\nu}+H^{^{(S)}}_{\mu\nu}+H^{^{(W)}}_{\mu\nu}
+H^{^{(SW)}}_{\mu\nu}\ ,
\end{equation}
where $S(W)$ labels the vector polarization of the target (spin of
the detected proton). All the effects caused by the first three
terms on the right side of Eq.~(47) were considered in previous
Sections and now we will investigate the radiative corrections to
the hadron double--spin correlations which arise just due to the
last term
$$H^{^{(SW)}}_{\mu\nu} =
(Sp_2)(Wp_1)\bigl[f_1\tilde g_{\mu\nu}+f_2\tilde
p_{1\mu}\tilde p_{1\nu}+f_3\tilde p_{2\mu}\tilde p_{2\nu} +f_4(\tilde p_1
\tilde p_2)_{\mu\nu}+if_5[\tilde p_1\tilde p_2]_{\mu\nu}\bigr] + $$
$$(Sp_2)(Wq)\bigl[f_6\tilde g_{\mu\nu}+f_7\tilde
p_{1\mu}\tilde p_{1\nu}+f_8\tilde p_{2\mu}\tilde p_{2\nu} +f_9(\tilde p_1
\tilde p_2)_{\mu\nu}+if_{10}[\tilde p_1\tilde p_2]_{\mu\nu}\bigr] + $$
$$(Sp_2)(WN)\bigl[f_{11}(\tilde p_1N)_{\mu\nu}+if_{12}[\tilde p_1N]_{\mu\nu}
+f_{13}(\tilde p_2N)_{\mu\nu}+if_{14}[\tilde p_2N]_{\mu\nu}\bigr]+ $$
$$(Sq)(Wp_1)\bigl[f_{15}\tilde g_{\mu\nu}+f_{16}\tilde p_{1\mu}\tilde
p_{1\nu} +f_{17}\tilde p_{2\mu}\tilde p_{2\nu} +f_{18}(\tilde p_1 \tilde
p_2)_{\mu\nu}+if_{19}[\tilde p_1\tilde p_2]_{\mu\nu}\bigr] + $$
\begin{equation}\label{48}
(Sq)(Wq)\bigl[f_{20}\tilde g_{\mu\nu}+f_{21}\tilde
p_{1\mu}\tilde p_{1\nu}+f_{22}\tilde p_{2\mu}\tilde p_{2\nu}
+f_{23}(\tilde p_1 \tilde p_2)_{\mu\nu}+if_{24}[\tilde p_1\tilde
p_2]_{\mu\nu}\bigr] + \end{equation}
$$(Sq)(WN)\bigl[f_{25}(\tilde p_1N)_{\mu\nu}+if_{26}[\tilde p_1N]_{\mu\nu}
+f_{27}(\tilde p_2N)_{\mu\nu}+if_{28}[\tilde p_2N]_{\mu\nu}\bigr]+ $$
$$(SN)(Wp_1)\bigl[f_{29}(\tilde p_1N)_{\mu\nu}+if_{30}[\tilde
p_1N]_{\mu\nu} +f_{31}(\tilde p_2N)_{\mu\nu}+if_{32}[\tilde
p_2N]_{\mu\nu}\bigr]+ $$
$$(SN)(Wq)\bigl[f_{33}(\tilde p_1N)_{\mu\nu}+if_{34}[\tilde p_1N]_{\mu\nu}
+f_{35}(\tilde p_2N)_{\mu\nu}+if_{36}[\tilde p_2N]_{\mu\nu}\bigr]+ $$
$$(SN)(WN)\bigl[f_{37}\tilde g_{\mu\nu}+f_{38}\tilde
p_{1\mu}\tilde p_{1\nu}+f_{39}\tilde p_{2\mu}\tilde p_{2\nu}
+f_{40}(\tilde p_1 \tilde p_2)_{\mu\nu}+if_{41}[\tilde p_1\tilde
p_2]_{\mu\nu}\bigr] \ . $$ Thus, the coefficients of the
polarization transfer from the target to the detected proton are
described, in general, by 41 structure functions. If the electron
beam is unpolarized, then the symmetrical part of the hadronic
tensor contributes only, and this corresponds to double--spin
(hadron--hadron) correlations in the cross--section of the process
(46). The antisymmetric part of the hadron tensor contributes in
the case of longitudinally--polarized electron beam due to
triple--spin (electron--hadron--hadron) correlations.

The corresponding radiatively corrected parts of the cross--section
for the unstable orientations of the target nucleus polarization
$S^{^J}$ (given by Eq.~(35)) and detected proton spin $W^{^I}$ (given
by Eq.~(17)) can be written as follows
\begin{equation}\label{49}
\varepsilon_2
E_2\frac{d\sigma_{JI}}{d^3k_2d^3p_2}=\sum_{j,i}A_{Jj}B_{Ii}\int\int
\frac{dx_1dx_2}{x_2^2}D(x_1)D(x_2)\hat\varepsilon_2
E_2\frac{d\hat{\sigma}^{^B}_{ji}}{d^3\hat k_2d^3p_2}
\end{equation}
where the Born cross--section under integral sign is defined
for the stable orientations of $S^{^j}$ (given by Eqs.~(33), (34)) and
$W^{^i}$ (given by Eqs.~(14), (15)) and depends on the $shifted$ variables
$$ \hat\varepsilon_2 E_2\frac{d\hat{\sigma}^{^B}_{ji}}{d^3\hat k_2d^3p_2}
=\hat\varepsilon_2 E_2\frac{d\sigma^{B}(\lambda,S^{j},W^{i},\hat
k_1,\hat k_2,p_1,p_2)}{d^3\hat k_2d^3p_2}\ . $$

In accordance with the calculations in Sections 3 and 4, matrices $A_{Jj}$
and $B_{Ii}$ are \begin{equation}\label{50}
A_{Jj}=\left(\begin{array}{ccc}1&0&0\\0&\cos{\theta}&-\sin{\theta}\\
0&\sin{\theta}&\cos{\theta}\end{array}\right), \ \
B_{Ii}=\left(\begin{array}{ccc}\cos{\theta_1}&\sin{\theta_1}&0\\
-\sin{\theta_1}&\cos{\theta_1}&0\\0&0&1\end{array}\right)\ ,
\end{equation}
$$I,J = L,T,N, \ \ i,j = l,t,n\ . $$

If we will write the hadron--hadron spin correlations in the Born
cross--section as
\begin{equation}\label{51}
\varepsilon_2E_2\frac{d\sigma^{^B}_{ji}}{d^2k_2d^3p_2}=\frac{\alpha^2
V^4X_{ji}}{16(2\pi)^32(2S_A+1)q^4}\ ,
\end{equation}
then the quantities $X_{ji}$ can be written in the form
\begin{equation}\label{52}
X_{ll}=2\sqrt{\frac{f\tau_1}{\tau_2}}\bigl\{\eta^2(R_{29}+\xi R_{33})+
\frac{2}{V^2\tau_1}\bigl[b(F_1+\xi F_6)-d(F_{15}+\xi F_{20})\bigr]\bigr\}\
,\end{equation}
\begin{equation}\label{53}
X_{lt}=\eta^2\sqrt{\frac{f}{\tau_1[1]}}\bigl[bR_{11}-dR_{25}+2\tau_1F_{37}
-\frac{2\psi}{\eta^2V^2f}\sqrt{[1]}(2bF_6-2dF_{20}+\eta^2V^2\tau_1R_{33})
\bigr]\ ,
\end{equation}
\begin{equation}\label{54}
X_{ln}=\frac{\eta}{\sqrt{\tau_1}}\bigl[\psi(bR_{11}-dR_{25}+2\tau_1F_{37})
+\frac{2}{V^2\sqrt{[1]}}(2bF_6-2dF_{20}+\eta^2V^2\tau_1R_{33})
\bigr]\ ,
\end{equation}
\begin{equation}\label{55}
X_{tl}=\sqrt{\frac{f}{r\tau_2}}\bigl\{\eta^2d(R_{29}+\xi R_{33})+\frac{4}
{V^2}\bigl[2r(F_{15}+\xi F_{20})+\zeta(F_1+\xi F_6)\bigr]\bigr\}\ ,
\end{equation}
\begin{equation}\label{56}
X_{tt}=\eta^2\sqrt{\frac{f}{r[1]}}\bigl[\zeta R_{11}+2rR_{25}+dF_{37}-
\frac{\psi\sqrt{[1]}}{\eta^2V^2f}(\eta^2V^2dR_{33}+4\zeta
F_6+8rF_{20})\bigr]\ ,
\end{equation}
\begin{equation}\label{57}
X_{tn}=\frac{\eta}{\sqrt{r}}\bigl[\psi(\zeta R_{11}+2rR_{25}+dF_{37})+
\frac{1}{V^2\sqrt{[1]}}(\eta^2V^2dR_{33}+4\zeta F_6+8rF_{20})\bigr]\ ,
\end{equation}
\begin{equation}\label{58}
X_{nl}=\eta\sqrt{\frac{f}{r\tau_2}}\bigl[\eta_1(R_{29}+\xi
R_{33})-\frac{4}{V^2}(F_1+\xi F_6)\bigr]\ ,
\end{equation}
\begin{equation}\label{59}
X_{nt}=\frac{\eta}{\sqrt{fr}}\bigl[\psi\bigl(\frac{4}{V^2}F_6 -\eta_1R_{33}
\bigr)+\frac{f}{\sqrt{[1]}}(\eta_1F_{37}-\eta^2R_{11})\bigr] \ ,
\end{equation}
\begin{equation}\label{60}
X_{nn}=-\frac{\eta^2}{\sqrt{r}}\bigl[\frac{1}{\sqrt{[1]}}\bigl(\frac{4}{V^2}
F_6 -\eta_1R_{33}
\bigr)-\psi(\frac{\eta_1}{\eta^2}F_{37}-R_{11})\bigr] \ .
\end{equation}
Here we used the following short notation
$$b=2z_1\tau_1-z, \ d=y(1+2x\tau_1), \ f=z^2-4\tau_1\tau_2, \
r=xy(1-y-xy\tau_1), \ $$
$$\zeta=z_2-z_1(1-y-2xy\tau_1)-xyz, \ \ \xi = \frac{z(z_1-z_2)-2y\tau_2}
{z^2-4\tau_1\tau_2}\ . $$

Functions $R_l$ and $F_l,$ which enter in the expressions for $X_{ji},$ are
defined by means of the hadron structure function $f$'s in Eq.~(48) as
\begin{equation}\label{61}
R_l=(2-y)f_l+(z_1+z_2)f_{l+2}+\frac{\lambda}{\eta}\bigl(\eta_1f_{l+1}
+\eta_2f_{l+3}\bigr) \ ,
\end{equation}
\begin{equation}\label{62}
F_l=-\frac{2xy}{V}f_l+(1-y-xy\tau_1)f_{l+1}+(z_1z_2-xy\tau_2)f_{l+2}+
(z_2+z_1(1-y)-xyz)f_{l+3}-\lambda\eta f_{l+4}\ .
\end{equation}
\section{Hadronic variables}

\hspace{0.7cm}

There exist the experimental possibility to measure the total
4--momentum of the hadron system $X$ instead to record the
scattered electron in semi--inclusive reactions. In such
experiments the momentum $q_h$ of heavy intermediate photon, that probes
the hadron structure, can be determined explicitly. The corresponding set
of dynamical variables is labeled usually as hadronic one.

In the case of the hadronic variables we have to eliminate the phase
space of the scattered electron and introduce the heavy photon phase space
by using the identity \begin{equation}\label{63}
\frac{d^3k_2}{\varepsilon_2}=2x_2^2x_h\frac{d^4q_h}{Q^2_h}\delta(x_1-x_h)
, \ \ \frac{d^4q_h}{Q_h^2}=\frac{dQ_h^2dx_hdy_hdz_h}{4x_h^2|\eta_h|}\
, \end{equation} $$x_h=-\frac{Q_h^2}{2k_1q_h}\ , \ \
y_h=\frac{2p_1q_h}{V}\ , \ \ z_h=\frac{2p_2q_h}{V}\ , \ \ Q_h^2 =-q_h^2 \
, $$
$$\eta_h^2=\frac{Q_h^2}{V}\Bigl[(4\tau_1\tau_2-z^2)\frac{Q_h^2}{x_h^2V}
+2\bigl(1-\frac{y_h}{x_h}\bigr)(zz_1-2\tau_2)+2\bigl(z_1-\frac{z_h}{x_h}\bigr)
(z-2z_1\tau_1)\Bigr]-(z_h-z_1y_h)^2\ .$$

Therefore, by combining representation (3) for the leptonic tensor and
(63), as well as bearing in mind the independence of the hadronic
tensor on variable $x_2,$ the expression for the quantity
$L_{\mu\nu}d^3k_2/\varepsilon_2,$ in the case of the hadronic variables
can be written as follows
\begin{equation}\label{64}
\frac{d^3k_2}{\varepsilon_2}L_{\mu\nu}=\frac{D(x_h,Q_h^2)}{x_h^2}
L^B_{\mu\nu}(\hat k_1,\hat k_1-q_h,\lambda)\frac{dx_h dy_h dz_h dQ_h^2}
{2|\eta_h|}\ .
\end{equation}
Note that for the events with undetected scattered electron the lower
limit of the integration over $x_2$ in Eq.~(3) equals to 0. In accordance
with the Kinoshita--Lee--Nauenberg theorem \cite{KLN}, the mass
singularities caused by the final--state radiation would disappear
in this case. On the language of the electron structure functions
this fact exhibits itself due to relation
$$\int\limits_0^1D(x,Q^2)dx=1\ ,$$ which was used to write Eq.~(64).

The lepton tensor in the Born approximation can be rewritten as
\begin{equation}\label{65}
L^B_{\mu\nu}(k_1,k_1-q_h)=2(k_1q_h)\tilde g_{\mu\nu}+4\tilde
k_{1\mu}\tilde k_{1\nu} -2i\lambda(\mu\nu k_1q_h)\ ,
\end{equation}
and the physically--founded parameterizations for $S^j$ in the process (1)
and $P^j$ in the process (32) remain now stable with respect to the scale
transformation $k_1\rightarrow x_hk_1.$ For example, one set can be
chosen as given by Eqs.~(14), (15) and other as $$S^L_{h\mu}=S^l_{\mu}, \
S^T_{h\mu}=\frac{(z^2-4\tau_1\tau_2)q_{h\mu}
+(2z_h\tau_1-zy_h)p_{2\mu}+(2y_h\tau_2-zz_h)p_{1\mu}}{\sqrt{V(z^2-4\tau_1
\tau_2)[q_h]}}, \ S^N_{h\mu}=\frac{2(\mu q_hp_1p_2)}{\sqrt{V^3[q_h]}}\ ,
$$
\begin{equation}\label{66}
[q_h]=
zz_hy_h+\frac{Q_h^2}{V}(z^2-4\tau_1\tau_2)-z_h^2\tau_1-y_h^2\tau_2, \ \
\end{equation}
with the transverse component in the plane $(\vec q_h, \vec p_2)$ in lab.
system.

Two physical sets of the target polarizations, both with the normal
component perpendicular to the plane $(\vec k_1, \vec q_h),$ may be chosen
as
\begin{equation}\label{67}
P^{^l}_{h\mu}=\frac{2\tau_1k_{1\mu}-p_{1\mu}}{M}, \ P^{^t}_{h\mu}=
\Bigl[-q_{h\mu}
+\bigl(y_h+\frac{2Q_h^2\tau_1}{x_hV}\bigr)k_{1\mu}-\frac{Q_h^2}{x_hV}
p_{1\mu}\Bigr]K^{-1}, \ P^{^n}_{h\mu}=\frac{-2(\mu k_1q_hp_1)}{VK}\ ,
\end{equation}
with the longitudinal component along $\vec k_1$ in lab. system and
$$P^{^L}_{h\mu}=\frac{2\tau_1q_{h\mu}-y_hp_{1\mu}}{MG}, \
P^{^T}_{h\mu}=\Bigl[\bigl(y_h^2+4\tau_1\frac{Q_h^2}{V}\bigr)k_{1\mu}-
\bigl(y_h+\frac{2Q_h^2\tau_1}{x_hV}\bigr)q_{h\mu}-\frac{Q_h^2}{V}\bigl(
2-\frac{y_h}{x_h}\bigr)p_{1\mu}\Bigr](KG)^{-1}\ , $$
\begin{equation}\label{68}
P^{^N}_{h\mu}=P^{^n}_{h\mu}, \ K=\sqrt{Q_h^2\bigl(1-\frac{y_h}{x_h}-
\frac{Q_h^2\tau_1}{x_h^2V}\bigr)}, \ G=\sqrt{y_h^2+4\frac{Q_h^2\tau_1}
{V}}\ ,
\end{equation}
with the longitudinal component along $\vec q_h.$ The different components
of the $P_h^{^J}$ in lab. system are
$$P_h^{^L}=\bigl(0, \ \vec n_q) , \ \ P_h^{^T} =
(0, \ \frac{\vec n_1-(\vec n_1\vec n_q)\vec n_q}{\sqrt{1-(\vec n_1\vec
n_q)^2}})\ , \ \ P_h^{^N}=(0, \ \frac{[\vec n_q\times\vec n_1]}{\sqrt{1-
(\vec n_1\vec n_q)^2}}) \ , $$
$$\vec n_q=\frac{\vec q_h}{|\vec q_h|}\ , \ \
\vec n_1 = \frac{\vec k_1}{|\vec k_1|} \ . $$

All these sets of proton spin and target polarization given by
Eqs.~(66), (67) and Eq.~(68), are stable with respect to th initial--state
collinear radiation. This can be verified by replacement $x_hk_1$
instead of $k_1$ at which
\begin{equation}\label{69}
k_1\rightarrow x_hk_1, \ x_h\rightarrow 1, \ y_h\rightarrow
\frac{y_h}{x_h}, \ z_h\rightarrow \frac{z_h}{x_h}, \ z\rightarrow
\frac{z}{x_h}, \ V\rightarrow x_hV, \ \tau_{1,2}\rightarrow
\frac{\tau_{1,2}}{x_h}\ .  \end{equation} To make the invariance
of $P^{^j} \ (j=l,t,n)$ and $P^{^J} (J=L,T,N)$ under replacement
(69) more transparent one can express $x_h$ in terms of $Q_h^2$
and $(k_1q_h).$ Then, for example,
$$ K=\sqrt{Q_h^2+y_h2(k_1q_h)-\frac{4(k_1q_h)^2\tau_1}{V}}, $$
and it is easy to see that this quantity is not changed under the
substitution (69).  Note also that quantity $\eta_h$ can be derived by
means of the rule $$\eta_h =x_h \eta^*,$$ where $\eta^*$ is determined
from $\eta$ with substitution $Q_h^2/V$ instead of $xy,$ $z_1-z_h$ instead
of $z_2$ and subsequent replacement (69).

That is why the cross--section for both the spin--independent and
spin--dependent parts in the case of the hadronic variables can be written
in the following form \begin{equation}\label{70}
E_2\frac{d\sigma^{^j}}{d^3p_2dQ_h^2dx_hdy_hdz_h}=\frac{D(x_h,Q_h^2)}{x_h^2}
E_2\frac{d\hat\sigma^{^B}_j}{d^3p_2dQ_h^2d\hat y_hd\hat z_h}\ ,
\end{equation}
where
$$E_2\frac{d\hat\sigma^{^B}_j}{d^3p_2dQ_h^2d\hat
y_hd\hat z_h}=\frac{\alpha^2C} {(2\pi)^3(2S_A+1)\hat V Q_h^4
2|\eta^*|}L_{\mu\nu}(\hat k_1,\hat k_1-q_h,
\lambda)H_{\mu\nu}(q_h,p_1,p_2;S^{^j}(P^{^j}))\ . $$
Here $C$ equals 1/2 (or 1) for process (1) or (32).
%To derive $\hat\eta_h$ it needs first to take $x_h=1$ in expression for
%$\eta_h$ and then to apply the replacement (69). Such operation leads to
%the relation $\eta_h=x_h\hat\eta_h,$ where multiplier $x_h$ is absorbed
%into definition of $\hat V.$

The representation (70) shows that the using of the hadron variables
allows to tag the initial--state radiated photon. Indeed, for fixed
4--momentum $P_x$ one can reconstruct 4--momentum $q_h$ and, consequently,
the variable $x_h$ which is the energy fraction of the photon radiated by
the initial electron (see Eq.~(63)).

The Born cross--section on the right side of Eq.~(70) has the form that is
very like to the corresponding cross--section for the leptonic variables.
We can formulate the following rules to write it: \\
i) change phase space differentials in the left sides of the expressions
valid for the leptonic variables
$$\frac{\varepsilon_2}{d^3k_2} \rightarrow \frac{2|\eta_{1h}|}
{dQ_h^2dy_hdz_h} \ , \ \ \eta_{1h} =\eta_h(x_h=1),$$
ii) apply substitution $$xy \rightarrow \frac{Q_h^2}{V}, \ y\rightarrow y_h, \
z_2\rightarrow z_1-z_h\ $$ to the right sides.

These rules lead, for example, to the formula for the
spin--dependent part of the cross--section of the process (1) in the case
of the longitudinal polarization (which follows from Eq.~(25))
\begin{equation}\label{71}
E_2\frac{d\sigma^{^B}_L}{d^3p_2dQ_h^2dy_hdz_h}=-\frac{\alpha^2V^3\eta_{1h}
\sqrt{z^2-4\tau_1\tau_2}}{8m(2S_A+1)(2\pi)^3Q_h^42|\eta_{1h}|}\Bigl[
H_2^{(h)}+\frac{zz_h-2y_h\tau_2}{z^2-4\tau_1\tau_2}H_3^{(h)}\Bigr]\ ,
\end{equation}
$$H_2^{(h)} =
(2-y_h)h_6+(2z_1-z_h)h_8+\frac{\lambda}{\eta_{1h}}(\eta_1^{^{(h)}}h_7
+\eta_2^{^{(h)}}h_9)\ , $$
$$\eta_1^{^{(h)}} = \frac{
Q_h^2}{V}[2(2z_1-z_h)\tau_1-z(2-y_h)]+z_1y_h^2-z_hy_h\ , \ \eta_2^{^{(h)}}=
\frac{Q_h^2}{V}[z(2z_1-z_h)-2(2-y_h)\tau_2]-z_h^2+z_1z_hy_h, $$
where $H_3^{(h)}$ is derived from $H_2^{(h)}$ by the change
$h_i\rightarrow h_{i+4}.$

The spin--dependent part of the cross--section of the process (32) for the
case of the normal target polarization (that follows from Eq.~(40)
reads \begin{equation}\label{72}
E_2\frac{d\sigma^{^B}_N}{d^3p_2dQ_h^2dy_hdz_h}=-\frac{\alpha^2V^3}{4(2S_A+1)
(2\pi)^3Q_h^4K(x_h=1)2|\eta_{1h}|}\Bigl\{\eta_{1h}^2G_2^{(h)}-\bigl[y_h(z_1y_h-z_h)+
\end{equation}
$$\frac{Q_h^2}{V}\bigl(2\tau_1(2z_1-z_h)-z(2-y_h)\bigr)\bigr]G_4^{(h)}\Bigr\}
\ .$$ The rest of the spin--dependent and spin--independent parts of the
cross--sections for processes (1) and (32) can be obtained by full analogy
using the above rules and results given in Sections 3, 4.

The variable $x_h$ characterizes the inelasticity of the
initial--state electron, and in the absence of radiation it equals
to 1. The electron structure function $D(x_h,Q_h^2)$ has
singularity at $x_h=1,$ and representation (70) shows that this
singularity is such that
\begin{equation}\label{73}
\lim\limits_{x_h\to 1} D(x_h,Q_h^2)dx_h = 1
\end{equation}
because in this limiting case the left side of Eq.~(70), being multiplied
by $dx_h,$ have to coincide with the Born cross--section.

\section{Conclusion}

\hspace{0.7cm}

In this paper we consider RC to the polarization observables in a wide
class
of semi--inclusive deep--inelastic processes.
We restrict ourselves to the leading--log accuracy and
neglect the contribution of the pair production in the singlet channel.
This gives the possibility to write the compact formulae for the
radiatively corrected spin--independent and spin--dependent parts of the
corresponding cross--sections in the form of the Drell--Yan representation
in electrodynamics by means of the electron structure functions.
The parameterization of the hadron spin 4--vectors in terms of the
particles 4--momenta is very important during the calculations. If
the momentum of the intermediate photon that probes the hadron
structure, is determined in terms of the hadronic variables, the traces of
the final--state radiation disappear in the final result in the framework
of used approximation.

In practice the corrections can be computed adopting some specific model
for structure functions. In this case the correction gets some model dependence
that can contribute to the systematical error in experimental
measurements. Another
way is related to some iteration procedure, when the fit of processed
experimental data is used for this required model. We note that obtained
leading
log formulae have a partly factorized form, being quite convenient for
this
procedure. The examples for DIS case can be found in \cite{P,POLRAD}.

Apart from the discussed classes of experiments the results can be also
adopted to exclusive electroproduction processes, when the unobservable
hadron
state is one particle.  In this case structure functions include an
additional $\delta$-function, so some analytical manipulations could be
necessary.

Sometimes the accuracy more than the leading one is necessary.  To
go beyond the leading accuracy one must modify the master
representations.  This modification concerns both the electron
structure function and cross--section (hard part) that depends on
the shifted variables. To improve the hard part, it is enough to
take into account the radiation of single additional
non--collinear photon and to add the non--leading part of the
one--loop correction. The corresponding procedure is described in
Ref.~\cite{KMF} for unpolarized deep--inelastic scattering and in
Ref.~\cite{P} (AAM) for quasi--elastic polarized electron--proton
scattering. To be complete one needs also to improve the
structure functions by the addition of the second order
next--to--leading contributions caused by double collinear photon
emission and pair production. Besides, the non--leading
contributions into $D$--function caused by the one--loop corrected
collinear single--photon emission and two--loop correction have to
be added properly. These contributions are different for
symmetric and asymmetric parts of the leptonic tensor and can be
extracted from the results given in Ref.~\cite{M} (for two--loop
correction, see \cite{BMR}). So, in this case we have to
distinguish between $D$ and $D_{\lambda}$ yet at the level of the
nonsinglet channel contribution. The concrete calculations will be
done elsewhere.

\end{document}